\documentclass[12pt]{article}
\usepackage{a4wide}
\usepackage{graphicx}
\def\1{\mbox{l\hspace{-0.53em}1}}
\DeclareMathAlphabet{\mathbbm}{U}{bbm}{m}{n}
  \SetMathAlphabet\mathbbm{bold}{U}{bbm}{bx}{n}

\begin{document}
\def\d{{\rm d}}
\title{\bf Mixed symmetric baryon multiplets  in large $N_c$ QCD: two and three flavours}
\author{N. Matagne
\thanks{{\it e-mail}: nicolas.matagne@umh.ac.be} ~and Fl.\ Stancu\thanks{{\it e-mail}: fstancu@ulg.ac.be} \\
 {\small $^\ast$ Service de Physique Nucl\'eaire et Subnucl\'eaire, University of Mons,}\\[-6pt]
 {\small Place du Parc, B-7000 Mons, Belgium,} \\ %
 {\small $^\dagger$ Institute of Physics, B5, University of Li\`ege,}\\[-6pt]
 {\small Sart Tilman, B-4000 Li\`ege 1, Belgium}}
\maketitle
\begin{abstract}\noindent
We propose a new method to study mixed symmetric multiplets of baryons in the context of 
the $1/N_c$ expansion approach. The simplicity of the method allows to better
understand the role of various operators acting on spin and flavour degrees of freedom. 
The method is tested on two and three flavours. It is shown that the spin and flavour operators proportional 
to the quadratic invariants of SU$_S$(2) and SU$_F$(3) respectively are dominant in the mass formula.
\end{abstract}

\section{Introduction}

The $1/N_c$ expansion method proposed by  't Hooft  \cite{HOOFT}
is a valuable tool to study nonperturbative dynamics  in a perturbative approach,
in terms of the parameter $1/N_c$  where $N_c$ is the number
of colors. 
The double line diagrammatic method of 
't Hooft implemented 
by Witten  \cite{WITTEN} to describe baryons gives 
convenient power counting rules for Feynman diagrams.
According to Witten's intuitive picture, a baryon containing $N_c$ quarks 
is seen as a bound 
state in an average self-consistent potential of a Hartree type 
and the corrections to the Hartree approximation are of order $1/N_c$.
These corrections capture the key phenomenological features of the baryon structure.
 


Ten years after 't Hooft's work, Gervais and Sakita  \cite{Gervais:1983wq}
and independently Dashen and Manohar in 1993  \cite{DM} discovered that QCD 
has an exact contracted SU(2$N_f$)$_c$ symmetry  
when $N_c \rightarrow \infty $,   $N_f$ being the number 
of flavors.
For ground state baryons the SU(2$N_f$) symmetry is broken by 
corrections proportional to $1/N_c$. 
Since 1993-1994 the $1/N_c$ expansion provided a systematic method to  
analyze baryon properties such as ground state masses, magnetic moments, axial currents, etc
\cite{Dashen:1994qi,Jenkins:1998wy,Jenkins:2001it,Jenkins:2009wm}.

A few years later the $1/N_c$ expansion method has been extended to excited states
also in the spirit of the Hartree approximation 
\cite{Goity:1996hk}. It was shown that for 
mixed symmetric states the SU($2N_f$) 
breaking occurs at order $N^0_c$ instead of $1/N_c$ as for the ground and symmetric
excited states.

Presently a lattice test of $1/N_c$ baryon masses relations has been performed \cite{Jenkins:2009wv}.
The lattice data clearly display both the $1/N_c$ and the SU(3) flavour symmetry breaking hierarchies.

Also, it was shown that the NN potential has an $1/N^2_c$ expansion 
and the strengths of the leading order central, spin-orbit, tensor and quadratic 
spin-orbit forces gave a qualitative understanding of the phenomenological meson 
exchange models \cite{Kaplan:1996rk}.


\section{The mass formula}\label{se:mf}

Here we are concerned with baryon spectra. The general form of the baryon mass operator is
\cite{Jenkins:1995td}
\begin{equation}
\label{massoperator}
M  = \sum_{i} c_i O_i + \sum_{i} d_i B_i
\end{equation} 
with the operators $O_i$  having the general form
\begin{equation}\label{OLFS}
O_i = \frac{1}{N^{n-1}_c} O^{(k)}_{\ell} \cdot O^{(k)}_{SF},
\end{equation}
where  $O^{(k)}_{\ell}$ is a $k$-rank tensor in O(3) and  $O^{(k)}_{SF}$
a $k$-rank tensor in SU(2), but invariant in SU($N_f$).
The latter is expressed in terms of  SU($N_f$) generators $S^i$, $T^a$ and $G^{ia}$ acting
on spin, flavour and spin-flavour respectively.
For the ground state one has $k$ = 0. 
Excited states require $k=1$ terms, which correspond to the angular
momentum component and the $k=2$ tensor term 
\begin{equation}\label{TENSOR}
L^{(2)ij}_{q}=\frac{1}{2}\left\{L^i_q,L^j_q\right\}
-\frac{1}{3}\delta_{i,-j}\vec{L}_q\cdot\vec{L}_q~.
\end{equation}

The first factor in  (\ref{OLFS})  gives the 
order $\mathcal{O}(1/N_c)$ of the operator in the series expansion
and reflects Witten's power counting rules.
The lower index $i$ represents a specific combination of generators, see
examples below. The $B_i$ are SU(3) breaking operators.
In the linear combination, Eq. (\ref{massoperator}),
$c_i$ and $d_i$ 
encode the QCD dynamics and are obtained from a fit to
the existing data. It is important to find regularities in their behaviour
\cite{Matagne:2005gd} and search for a possible compatibility with quark models
\cite{Semay:2007cv}.

A considerable amount of work has been devoted to ground state baryons
summarized in  several review  papers as, for example,
\cite{Dashen:1994qi,Jenkins:1998wy,Jenkins:2001it}.
The ground state is
described by the symmetric representation $[N_c]$. For $N_c$ = 3 
this becomes $[3]$ or  $[\bf 56]$ in an SU(6) dimensional notation.

In the following we shall concentrate on the description of excited states only and the motivation 
will be obvious.

\section{Excited states}\label{se:excit}

Excited baryons can be divided into SU(6) multiplets, as in the
constituent quark model.
If an excited baryon belongs to the $[\bf{56}]$-plet
the mass problem can be treated similarly to the ground state
in the flavour-spin degrees of freedom, but one has to take into
account the presence of an orbital excitation in the space
part of the wave function  \cite{Goity:2003ab,Matagne:2004pm}.
If the baryon belongs to 
the mixed symmetric representation $[21]$, or $[\bf{70}]$ in SU(6)
notation, the treatment becomes much more complicated. 

There is  a standard way to study the   $[\bf{70}]$-plets
which is related to the Hartree approximation  \cite{Goity:1996hk}.
An excited baryon is described by symmetric core plus 
an excited quark coupled to this core, see \emph{e.g.} 
\cite{Carlson:1998vx,Goity:2002pu,Matagne:2006zf,Matagne:2006wi}.
In that case the core 
can be treated in a way similar to that of the ground state.
In this method each SU($2N_f$) $\times$ O(3) generator is  splitted 
into two terms 
\begin{equation}\label{CORE}
S^i = s^i + S^i_c; ~~~~T^a = t^a + T^a_c; ~~~ G^{ia} = g^{ia} + G^{ia}_c,
~~~ \ell^i = \ell^i_q + \ell^i_c,
\end{equation}
where  $s^i$, $t^a$, $g^{ia}$ and $\ell^i_q$  are the excited 
quark operators and  
$S^i_c$, $T^a_c$, $G^{ia}_c$ and  $\ell^i_c$ the corresponding core operators.

In this procedure the wave function is approximated by the
term which corresponds  to the normal Young tableau, where
the decoupling of the excited quark is straightforward.
The other terms needed to construct a symmetric orbital-flavour-spin
state are neglected, \emph{i.e.} antisymmetry is ignored. 
An a posteriori  justification is given in Ref. \cite{Pirjol:2007ed}. 

But the number of  linearly
independent operators constructed from the generators given in the
right-hand side of Eqs. (\ref{CORE}) 
increases tremendously the number of terms in the mass formula so that the number of coefficients 
to be determined usually becomes much larger than the experimental data
available. Consequently, in selecting the most dominant operators 
one has to make an arbitrary choice, as for example in Ref.  \cite{Carlson:1998vx}.
In particular the isospin operator as $t \cdot T^c/N_c $ , although important, has been entirely ignored
without any reason.

A solution to this problem has been found in Ref.  \cite{Matagne:2006dj},
where the separation into a symmetric core and an excited quark is
not necessary.  The key issue  is the knowledge of the matrix elements of
the SU(2N$_f$) generators for mixed symmetric states described by the
partition $[N_c-1,1]$ for arbitrary $N_c$. These can be obtained  by using  
a generalized Wigner-Eckart  theorem \cite{Hecht:1969ck}.
Using  SU(2N$_f$) generators acting on the whole system, the number
of operators up to $1/N_c$ order in the mass formula is considerably reduced so that the 
physics becomes more transparent, as we shall see below.

\subsection{The SU(4) case}

The SU(4) case has been presented in Ref. \cite{Matagne:2006dj}. Its algebra is
\begin{eqnarray}\label{ALGEBRASU4}
&[S^i,S^j]  =  i \varepsilon^{ijk} S^k,
~~~~~ [T^a,T^b]  =  i \varepsilon^{abc} T^c,\nonumber \\
&[G^{ia},G^{jb}] = \frac{i}{4} \delta^{ij} \varepsilon^{abc} T^c
+\frac{i}{2} \delta^{ab}\varepsilon^{ijk}S^k,
\end{eqnarray}
with $i, a = 1,2,3$. 
The matrix elements of the SU(4) generators were extracted from Ref. \cite{Hecht:1969ck},
initially proposed for nuclear physics where SU(4) symmetry is nearly exact.
The transcription to a system of $N_c$ quarks was straightforward.
Instead of 12 operators up to order $\mathcal{O}(1/N_c)$ presented  in Ref.  \cite{Carlson:1998vx} we needed 
only 6 operators for 7 experimentally known three- and four-star nonstrange resonances (no mixing angles).
We have introduced the spin and isospin operators on equal footing, 
as seen from Table \ref{nonstrange}, and obtained the new result that
the isospin term $O_4$ becomes as dominant in $\Delta$ resonances 
as the spin term $O_3$ does in $N^*$ resonances,  as indicated by 
the comparable size of the coefficients $c_3$ and $ c_4$ in Table  \ref{nonstrange}.
Column 5 proves that by the removal of  $O_4$  the fit deteriorates considerably.

\begin{table*}[ht!]
\caption{List of operators $O_i$ and coefficients $c_i$ in the $N=1$ band revisited, 
7 resonances of  3 and 4 stars status, no mixing angles.}
\vspace{0.2cm}
\label{nonstrange}
{\scriptsize
\renewcommand{\arraystretch}{2} 
\begin{tabular}{lrrrrr}
\hline
\hline
Operator \hspace{2cm} &\hspace{0.0cm} Fit 1 (MeV) & \hspace{0.3cm} Fit 2 (MeV) & \hspace{0.3cm}Fit 3 (Mev) &\hspace{0.3cm} Fit 4 (MeV) &\hspace{0.5cm} Fit 5 (MeV) \\
\hline
$O_1 = N_c  \ \1$                            & $481 \pm5$  & $482\pm5$ &  $484\pm4$ &  $484\pm4$ & $498\pm3$ \\
$O_2 = \ell^i s^i$                	     & $-31 \pm26$ & $-20\pm23$ & $-12\pm20$ & $3\pm15$ & $38\pm34$ \\
$O_3 = \frac{1}{N_c}S^iS^i$                  & $161\pm 16$ & $149\pm11$ & $163\pm16$ & $150\pm11$  &$156\pm16$\\
$O_4 = \frac{1}{N_c}T^aT^a$                  & $169\pm36$  & $170\pm36$ & $141\pm27$ & $139\pm27$ \\
$O_5 = \frac{15}{N_c}L^{(2)ij}G^{ia}G^{ja}$    & $-29\pm31$&            & $-34\pm30$&        & $-34\pm31$ \\
$O_6 = \frac{3}{N_c}L^iT^aG^{ia}$            & $32\pm26$ & $35\pm26$ &            &      & $-67\pm30$ \\
\hline
$\chi_{\mathrm{dof}}^2$                                    & $0.43$      & $0.68$ & $0.94$           & $1.04$ & $11.5$ \\
\hline \hline
\end{tabular}}
\end{table*}
\subsection{The SU(6) case}

Below we present preliminary results for SU(6).
The group algebra is
\begin{eqnarray}\label{ALGEBRA}
&[S^i,S^j]  =  i \varepsilon^{ijk} S^k,
~~~~~[T^a,T^b]  =  i f^{abc} T^c, \nonumber \\
&[S^i,G^{ja}]  =  i \varepsilon^{ijk} G^{ka},
~~~~~[T^a,G^{jb}]  =  i f^{abc} G^{ic}, \nonumber \\
&[G^{ia},G^{jb}] = \frac{i}{4} \delta^{ij} f^{abc} T^c
+\frac{i}{2} \varepsilon^{ijk}\left(\frac{1}{3}\delta^{ab} S^k 
+d^{abc} G^{kc}\right),
\end{eqnarray}
with $i$ = 1,2,3
and $a$ = 1,2,...,8.
The analytic work was based on the extension  of Ref. \cite{Hecht:1969ck} from SU(4)  to SU(6)  in order to obtain 
matrix elements of all SU(6) generators between symmetric $[N_c]$ states  first \cite{Matagne:2006xx},  followed later
by  matrix elements of all SU(6) generators between mixed symmetric states $[N_c-1,1]$ states   \cite{Matagne:2008kb}.
The latter work has been recently completed by some new isoscalar factors  required by the physical problem  \cite{MSnew}.

Theoretically the $[{\bf 70},1^-]$ multiplet has 5 octets ($N, \Lambda, \Sigma, \Xi$), 2 decuplets ($\Delta, \Sigma, \Xi, \Omega$) 
and two flavour singlets $\Lambda_{1/2}$ and $\Lambda_{3/2}$.
In the fit we take into account  the 17 experimentally known resonances
having a 3 or 4 star status and the two known mixing angles between the $^2N_J$ and $^4N_J$ ($J$ = 1/2, 3/2) states.
Table \ref{operators} exhibits the  9 operators used in the mass formula,  from which the three $B_i$'s break 
explicitly the SU(3) symmetry.  The corresponding  fitted coefficients $c_i$ and $d_i$ are indicated under a preliminary fit
named Fit 1.  We remind that  in the symmetric core + excited quark procedure 
fifteen $O_i$ (flavour invariants) and four $B_i$ operators were included  in the fit \cite{SGS}.  However the flavour operator
$1/N_c ~t \cdot T^c$ was omitted, without any justification.

Like for nonstrange baryons,  one can see that the dominant operators are the spin $O_3$ and flavour $O_4$.
The latter has the form explained in Ref. \cite{Matagne:2008kb}. It recovers  the matrix elements of $O_4 = 1/N_c ~T^a T^a$  
of nonstrange baryons (see Table \ref{nonstrange}). The operators $O_3$ and $O_4$ have similar values for
the corresponding coefficients, which proves the importance of the flavour operators in the fit, like for the
SU(4) case. 

\begin{center}
\begin{table}[ht!]
\caption{Operators and their coefficients in the mass formula obtained from 
a numerical fit, mixing angles included, $\mathcal{S}$ denotes the strangeness.} 
\vspace{0.2cm}
\label{operators}{\scriptsize
\renewcommand{\arraystretch}{2} 
\begin{tabular}{lrrrr}
\hline
\hline
Operator \hspace{3cm} &\hspace{0.8cm} Fit 1 (MeV) 
\\
\hline
$O_1 = N_c \ \1$                                                            &  $476.11 \pm 4.09$ &      \\
$O_2 = l^i s^i$                	                                              &    $63.6 \pm 22.6$  &     \\
$O_3 = \frac{1}{N_c}S^iS^i$                                             &  $165 \pm 15$  &      \\
$O_4 = \frac{1}{N_c} (T^aT^a - \frac{1}{12} N_c(N_c+6))$  &  $181.95 \pm 11.6$ &      \\
$O_5 = \frac{3}{N_c} L^iT^aG^{ia}$                                   &   $-19.4 \pm 6$   &       \\
$O_6 = \frac{15}{N_c} L^{(2)ij}G^{ia}G^{ja}$                      &   $8.5 \pm 0.3$   &           \\
\hline
$B_1$ = - $\mathcal{S}$  &  $163.90 \pm 12.04$  &   \\
$B_2$= $\frac{1}{N_c} L^i G^{i8} -  \frac{1}{2} \sqrt{\frac{3}{2}}  O_2$    &  $33.96 \pm 31.55$   &   \\
$B_3$ = $\frac{1}{N_c} S^i G^{i8} - \frac{1}{2 \sqrt{3}} O_3 $    &   $112.46  \pm 62.14$  &   \\

\hline
$\chi_{\mathrm{dof}}^2$   &  2.85  &     \\
\hline \hline
\end{tabular}}
\end{table}
\end{center}
In Tables   \ref{nonstrange} and \ref{operators}  the operator  $O_2$ contains the one-body part of the spin-orbit term,
defined in Ref. \cite{Carlson:1998vx},
while $O_5$, ~$O_6$ and $B_2$ contain the total orbital angular momentum components $L^i$, as in Eq.   (\ref{TENSOR}).
 Using the 
total spin-orbit term it would hardly affect the fit.  The contribution of terms containing the angular 
momentum is generally small, like for nonstrange baryons \cite {Matagne:2006dj}, see Table \ref{nonstrange}. 
The SU(3) breaking operator $B_1$ turns out to be important, as expected.  

The $\chi_{\mathrm{dof}}^2 =  2.85$ is larger than desired. We found that the basic reason is
that it is hard to fit the mass
of $\Lambda (1405)$ to be so low.  The difficulty  is entirely similar to that of quark models,
where  $\Lambda (1405)$  appears too high.  An artificially larger mass of the order of 1500 MeV
considerably improves the fit, leading to $\chi_{\mathrm{dof}}^2 < 1$. More fits will be presented elsewhere \cite{MSnew}. 

The difference between our results and those of Ref. \cite{Goity:2002pu} can partly be explained as due 
to the difference in the wave function. In Ref. \cite{Goity:2002pu} only the component with 
$S_c$ = 0 is taken into account and this component brings no contribution to the
spin term in flavour singlets, so that the mass of  $\Lambda (1405)$ remains low.
In our case, where we use 
the exact wave function, both $S_c$ = 0 and  $S_c$ = 1 parts of the wave function 
contribute to the spin term. This makes the 
spin term contribution identical  for all states of given $J$ irrespective of the flavour, which seems to
us natural.  Then, in our case,  with a non vanishing spin term in flavour singlets as well,
the mass formula accomodates a heavier  $\Lambda (1405)$ than the experiment,
like in quark models (for a review on the controversial nature of $\Lambda (1405)$ see, for example,
Ref. \cite{Pakvasa:1999zv} where one of the authors S.F. Tuan has predicted together with 
D.H. Dalitz this resonance in 1959, discovered experimentally two years later.)


\section{Conclusion}\label{se:concl}

The $1/N_c$ expansion method provides a powerful theoretical
tool to analyze the spin-flavour symmetry of baryons and
explains the success of models based on this symmetry.
We have shown that the dominant contributions come from the spin and flavour terms in 
the mass formula both in SU(4) and SU(6).  The terms containing angular 
momentum bring small contributions, which however slightly improve the 
fit. It is hard to fit the mass
of  $\Lambda (1405)$,  a notorious problem in realistic  quark models
\cite{Capstick:1986bm,Glozman:1997ag}.
This  suggests again a 
more complex nature of this resonance, as, for example,  a coupling to a $\bar K N$ system ,
which might survive in the large $N_c$ limit \cite{GarciaRecio:2006wb,Hyodo:2007np}.

 
\end{document}